\documentstyle[aps]{revtex}

\def\nad#1{\mbox{\smash{\oalign{$#1$\crcr\hidewidth$\mathchar"017E$
\hidewidth}}}}
\def\Nad#1{\mbox{\smash{\oalign{$#1$\crcr\hidewidth$\,\mathchar"707E$
\hidewidth}}}}
\def\ltimes{\,\rule{.4pt}{4pt}\!\!\times}

\advance\oddsidemargin by -12pt \advance\evensidemargin by -12pt
\advance\topmargin by .34in

\begin{document}

\title{\marginpar{\vspace{-.5in}\hspace{-1in}\small KFT U{\L} 5/94}
Tachyonic neutrinos?}
\author{Jakub Rembieli\'nski\thanks{This work is supported under the
{\L}\'od\'z University grant no.\ 457.}}
\address{Katedra Fizyki Teoretycznej, Uniwersytet {\L}\'odzki\\
ul.~Pomorska 149/153, 90--236 {\L}\'od\'z, Poland\/%
\thanks{{\it E-mail address\/}: jaremb@mvii.uni.lodz.pl}}
\maketitle

\begin{abstract}
It is shown that tachyons are associated with unitary
representations of Poincar\'e mappings induced from $SO(2)$
little group insted of $SO(2,1)$ one. This allows us to treat
more seriously possibility that neutrinos are fermionic
tachyons according to the present experimental data.
\end{abstract}


\section{Introduction}
Almost all recent experiments, measuring directly or indirectly the
electron and muon neutrino masses, have yielded negative values for the
mass square \cite{tab}. It suggests that these particles might be
fermionic tachyons. This intriguing possibility was written down some
years ago by Chodos {\em et al.\/} \cite{CHK}.  Furthermore, possible new
experiments are presented in the papers by Kostelecky \cite{Kos1} and
Chodos {\em et al} \cite{CKPG,CK}.

On the other hand, in the current opinion, there is no
satisfactory theory of superluminal particles.
This persuasion creates a psychological barrier to take such
possibility seriously. Even if we consider eventuality that
neutrinos are tachyons, the next problem arises; namely a
modification of the theory of electro-weak interaction will be
necessary in such a case. But, as we known, in the standard
formulation of special relativity, the unitary representations
of the Poincar\'e group, describing fermionic tachyons, are
induced from infinite dimensional unitary representations of the
noncompact $SO(2,1)$ little group.  Consequently, the neutrino
field should be infinite-component one so a construction of an
acceptable local interaction is extremally difficult.

In this paper we suggest a solution to the above dilemma. To do
this we use the formalism developed in the paper \cite{Rem3}
based on the earlier works \cite{Rem1,Rem2}, where it was
proposed a consistent description of tachyons on both classical
and quantum level. The basic idea is to extend the notion of
causality without a change of special relativity. This can be
done by means of a freedom in the determination of the notion of
the one-way light velocity, known as the ``conventionality
thesis'' \cite{Rei,Jam}. The main results obtained in
\cite{Rem3} can be summarized as follows:
\begin{itemize}
\item
The relativity principle is formulated in the framework of a nonstandard
synchronization scheme (the Chang--Tangherlini (CT) scheme).  The absolute
causality holds for all kinds of events (time-like, light-like, space-like).
\item
For bradyons and luxons our scheme is fully equivalent to the standard
formulation of special relativity.
\item
For tachyons it is possible to formulate covariantly  proper initial
conditions.
\item
There exists a covariant lower bound of energy for tachyons.
\item
The paradox of ``transcendental'' tachyons is solved.
\item
Tachyonic field can be consistently quantized using the CT synchronization
scheme.
\item
Tachyons distinguish a preferred frame {\em via\/} mechanism of
the relativity principle breaking \cite{Rem2,Rem3}.
\end{itemize}
In this paper we make the next step in this direction by classification
of all possible unitary Poincar\'e mappings for space-like momenta. The
main and unexpected result of the present work is that unitary mappings
for space-like momenta are induced from the $SO(2)$ little group. This
holds because we have a bundle of Hilbert spaces rather than a single
Hilbert space of states. Therefore unitary operators representing
Poincar\'e group act in irreducible orbits in this bundle. Consequently,
elementary states are labelled by helicity, in an analogy with the
light-like case. This fact is extremally important because we have no
problem with infinite component field.

Now, let us begin with
a brief review of the theory proposed in \cite{Rem3,Rem1,Rem2}.

It is rather evident that a consistent description of tachyons
lies in a proper extension of the causality principle. Note that
interpretation of the space-like world lines as physically
admissible tachyonic trajectories favour the constant-time
initial hyperplanes. It follows from the fact that only such
surfaces intersect each world line with locally nonvanishing
slope once and only once.  Notwithstanding, the instant-time
hyperplane is not a Lorentz-covariant notion which is
just the source of many troubles with causality. In the standard
framework of the special relativity, space-like geodesics do not
have their physical counterparts. This is an immediate
consequence of the assumed causality principle which admits
time-like and light-like trajectories only.

It is important to stress the following two well known facts
from special relativity:
\begin{itemize}
\item
The definition of a coordinate time
depends on the synchronization scheme \cite{Rei,VT,Var},
\item
Synchronization scheme is a convention, because no
experimental procedure exists which makes it possible to
determine the one-way velocity of light without use of
superluminal signals \cite{Jam}.
\end{itemize}
Therefore a choice of a synchronization scheme does not affect
the assumptions of special relativity but evidently it can
change the notion of causality, depending on the definition of
the coordinate time.

Following Einstein, intrasystemic synchronization of clocks in
their ``setting'' (zero) requires a definitional or conventional
stipulation (for discussion see Jammer \cite{Jam} and Sj\"odin
\cite{Sjo}).  Indeed, to determine one-way light speed it is
necessary to use synchronized clocks at rest in their
``setting'' (zero). On the other hand to synchronize clocks we
should know the one-way light velocity.  Thus we have a logical
loophole.  Therefore no experimental procedure exists (if we
exclude superluminal signals) which makes possible to determine
unambigously and without any convention the one-way velocity of
light. Consequently, only the average value of the light
velocity around closed paths has an operational meaning. This
statement is known as the conventionality thesis \cite{Jam}.
However, the requirement of causality, logically independent on
the requirement of the Lorentz covariance, can contradict the
conventionality thesis and  consequently it can prefer a
definite synchronization scheme, namely CT scheme.

In the papers by Chang \cite{Cha1,Cha2,Cha3}, it was introduced
four-dimensional version of the Tangherlini transformations
\cite{Tan}, termed the Generalized Galilean Transformations
(GGT).  In \cite{Rem1} it was shown that GGT, extended to form a
group, are nonlinear form of the Lorentz group transformations
with $SO(3)$ as a stability subgroup.  The coordinate
transformations should be supplemented by transformations of a
vector-parameter interpreted as the velocity of a privileged
frame.  It was also shown \cite{Rem1}  that the above family of
frames is equivalent to the Einstein--Lorentz one. A difference
lies in another synchronization procedure for clocks.  As a
consequence a constant-time hyperplane is a covariant notion in
our formalism.  Hereafter we call this procedure of
synchronization the Chang--Tangherlini synchronization scheme.

In the papers \cite{Rem2,Rem3} these ideas was developed and
applied to the description of tachyons.

\section{Formalism}
Let us start with a simple observation that the description of a
family of relativistic inertial frames in the Minkowski
space-time is not so natural.  Instead, it seems
that the geometrical notion of bundle of frames is much more natural.
Base space is identified with the space of velocities; each
velocity marks out a coordinate frame.  Indeed, from the point
of view of an observer (in a fixed inertial frame), all inertial
frames are labelled by their velocities with respect to him.
Therefore, in principle, to define the transformation rules
between frames, we can use, except of coordinates, also this
vector parameter, possibly related to velocities of frames with
respect to a distinguished observer.

According to the paper \cite{Rem3}, transformation between two
coordinate frames $x^{\mu}$ and ${x'}^\mu$ has the following
form
\begin{mathletters}\label{1}
\begin{equation}
x'=D(\Lambda,u)(x+a),
\end{equation}
\begin{equation}
u'=D(\Lambda,u)u.
\end{equation}
\end{mathletters}
Here $\Lambda$ belongs to the Lorentz group $L$, whilst $u$ is a
fourvelocity of a privileged inertial frame measured in the coordinate
frame $x^{\mu}$\footnote{A necessity of a presence of a preferred frame
for tachyons was stressed by many authors (see, for example,
\cite{CKPG,Sud}).}. The $a^{\mu}$ are translations. The transformations
(\ref{1}) have standard form for rotations i.e. $D(R,u)=R$, whereas for
boosts the matrix $D$ takes the form
\begin{equation}
D(\vec{V},\vec{\sigma})=\left(\begin{array}{c|c} \gamma & 0\\[1ex]
\hline \displaystyle-\frac{\vec V}{c}\gamma^{-1} & I+
\displaystyle\frac{\vec V\otimes\vec V^{\rm T}}
{c^2\gamma\left[\gamma+\sqrt{\gamma^2+\displaystyle\frac{\vec V}{c}^2}
\right]}-\frac{\vec V\otimes\vec\sigma^{\rm T}}{c^2\gamma\gamma_0}
\end{array}\right)                                             \label{2}
\end{equation}
where we have used the following notation
\begin{mathletters} \label{3}
\begin{equation}
\gamma_0=\left[\frac{1}{2}\left(1+\sqrt{1+\left(\displaystyle
\frac{2\vec\sigma}{c}\right)^2}\right)\right]^{1/2},
\end{equation}
\begin{equation}
\gamma(\vec V)= \left(\left(1+\frac{\vec\sigma\vec V}{c^2}
\gamma^{-2}_0\right)^2 -\left(\frac{\vec
V}{c}\right)^{2}\right)^{1/2},
\end{equation}
\begin{equation}
\frac{\vec{\sigma}}{c}=\frac{\vec{u}}{u^0}.
\end{equation}
\end{mathletters}
Here $\vec{V}$ is a relative velocity of $x'$ frame with respect to $x$
whilst $\vec{\sigma}$ is the velocity of a preferred frame. The
transformations (\ref{2}) remain unaffected the line element
\begin{equation}
ds^2=g_{\mu\nu}(u)dx^{\mu}dx^{\nu}                          \label{4}
\end{equation}
with
\begin{equation} g(u)=
\left(\begin{array}{c|c}
 1 & \displaystyle\frac{\vec\sigma^{\rm T}}{c}\gamma_0^{-2}\\[1ex]
\hline
\displaystyle\frac{\vec\sigma}{c}\gamma_0^{-2}&
\displaystyle-I+\frac{\vec\sigma\otimes\vec\sigma^{\rm T}}{c^{2}}
\gamma_0^{-4}
\end{array}\right),                                 \label{5}
\end{equation}
Notice that $u^2=g_{\mu\nu}(u)u^{\mu}u^{\nu}=c^2$.

{}From (\ref{4}) we can calculate the velocity of light propagating
in a direction $\vec{n}$
\begin{equation}
\vec c=\frac{c\vec n}{1-\displaystyle\frac{\vec n\vec\sigma}{c}
\gamma_0^{-2}}.                                                   \label{6}
\end{equation}
It is easy to verify that the average value of $\vec{c}$ over a
closed path is always equal to $c$.

Now, according to our interpretation of the freedom in
realization of the Lorentz group as freedom of the
synchronization convention, there should exists a relationship
between $x^{\mu}$ coordinates and the Einstein-Poincar\'e (EP)
ones denoted by $x^{\mu}_{E}$. Indeed, we observe, that the
coordinates
\begin{mathletters}\label{7}
\begin{equation}
x_E=T^{-1}(u)x,\end{equation}\begin{equation}
u_E=T^{-1}(u)u,
\end{equation}
\end{mathletters}
where the matrix $T$ is given by
\[
T(\vec\sigma)=\left(\begin{array}{c|c}
1&\displaystyle-\frac{\vec\sigma^{\rm T}}{c}\gamma_0^{-2}\\[1ex]
\hline
0&I\end{array}\right).
\]
transform under the Lorentz group standardly i.e. (\ref{1}) and
(\ref{7}) imply
\begin{mathletters}\label{8}
\begin{equation}
x^{\prime}_{E}=\Lambda x_E,\end{equation}\begin{equation}
u^{\prime}_{E}=\Lambda u_E.
\end{equation}
\end{mathletters}
It holds because $D(\Lambda, u) = T(u') \Lambda T^{-1}(u)$.
Moreover, $ds^2=ds^{2}_{E},\quad \vec{c}_E=c \vec{n},\quad
u_{E}^{2}=c^2$ and $g_E={\rm diag}(+,-,-,-)$. Therefore the
CT synchronization scheme defined by the
transformations rules (\ref{1}) is at first glance equivalent to
the EP one. In fact, it is a different choice of
the convention of the one-way light propagation (see (\ref{6}));
in other words it is another formulation of special relativity.
Notwithstanding, the above statement is true only if we exclude
superluminal signals. Indeed, the causality principle, logically
independent of the requirement of Lorentz covariance, is not
invariant under change of the synchronization (\ref{7}). It is
evident from the form of the boost matrix (\ref{2}); the
coordinate time $x^0$ is rescaled by a positive factor $\gamma$
only. Therefore $\varepsilon(dx^0)$  is an invariant of
(\ref{1}) and this factor allow us to introduce an absolute
notion of causality, generalizing the EP causality.
Consequently, as was shown in \cite{Rem3}, all inconsistencies
of the standard formalizm, related to the superluminal
propagation, disappear in this formulation of special
relativity.

If we exclude tachyons then, as was mentioned above, physics
cannot depend of synchronization. Thus in this case {\em any
inertial frame can be choosen as the preferred frame \/},
determining a concrete CT synchronization. This statement is in
fact the relativity principle articulated in the CT
synchronization language.

What happens, when tachyons do exist? In such a case
the relativity principle is obiously broken: If tachyons exist
then only one inertial frame must be a {\em true privileged frame}.
Therefore, in this case, the EP synchronization is unadequate to
description of reality; we must choose the synchronization
defined by (\ref{1}--\ref{6}).  Moreover the relativity
principle is evidently broken in this case as well as the
conventionality thesis: The one-way velocity of light becomes
({\em a priori}) a really measured quantity.

To formalize the above analysis, in \cite{Rem3} it was introduced
notion of the synchronization group $L_S$. It connects different
synchronizations of the CT--type and it is isomorphic to the
Lorentz group:
\begin{mathletters} \label{9}
\begin{equation}
x'=T(u')T^{-1}(u)x=D(\Lambda_S,u)T(u)\Lambda_{S}^{-1}T^{-1}(u)x,
\end{equation}\begin{equation}
u'=D(\Lambda_S,u)u,
\end{equation}
\end{mathletters}
with $\Lambda_S\in L_S$.

For clarity we write the composition of transformations of the
Poincar\'e group $L\ltimes T^4$ and the synchronization group $L_S$ in
the EP coordinates
\begin{mathletters} \label{10}
\begin{equation}
x_{E}^{\prime}=\Lambda(x_E+a_E),\end{equation}\begin{equation}
u_{E}^{\prime}=\Lambda_S\Lambda u_E.
\end{equation}
\end{mathletters}
Therefore, in a natural way, we can select three subgroups:
\[
L=\{(I,\Lambda)\},\quad L_S=\{(\Lambda_S,I)\},\quad
 L_0=\{(\Lambda_{0},\Lambda^{-1}_{0})\}.
\]
By means of (\ref{10}) it is easy to check that $L_0$ and $L_S$ commute.
Therefore the set $\{(\Lambda_S,\Lambda)\}$ is simply the direct product of
two Lorentz groups $L_0\otimes L_S$.  The intersystemic Lorentz group $L$ is
the diagonal subgroup in this direct product.  From the composition law
(\ref{10}) it follows that $L$ acts as an authomorphism group of $L_S$.

Now, the synchronization group realizes in fact the relativity
principle: If we {\em exclude tachyons} then transformations of
$L_S$ are canonical ones. On the other hand, if we {\em include
tachyons} then the synchronization group $L_S$ is broken to the
$SO(3)_u$ subgroup of $L_S$; $SO(3)_u$ is the stability group of
$u^{\mu}$. In fact, transformations from the $L_S/SO(3)_u$ do
not leave the absolute notion of causality invariant. On the
quantum level $L_S$ is broken
down to $SO(3)_u$ subgroup i.e. transformations from $L_S/SO(3)_u$
cannot be realized by unitary operators \cite{Rem3}.


\section{Quantization}
The following two facts, true only in CT synchronization, are
extremally important for quantization of tachyons:
\begin{itemize}
\item
Invariance of the sign of the time component of the space-like
fourmomentum i.e. $\varepsilon(k^0)=\mbox{\em inv}$,
\item
Existence of a covariant lower energy bound.
\end{itemize}
This is the reason why an invariant Fock construction can be
done in our case \cite{Rem3}.  In the paper \cite{Rem3} it was constructed a
quantum free field theory for scalar tachyons. Here we classify
unitary Poincar\'e mappings in the bundle of Hilbert spaces
$H_u$ for a space-like fourmomentum. Furthermore we find the
corresponding canonical commutation relations. As result we
obtain that tachyons correspond to unitary mappings which are
induced from $SO(2)$ group rather than $SO(2,1)$ one.
Of course, a classification of unitary representations for
time-like and light-like fourmomentum is the same as in EP
synchronization; this holds because the relativity principle is
working in this case.

\subsection{Tachyonic representations}
As usually, we assume that a basis in a Hilbert space  $H_u$
(fibre) of one-particle states consists of
the eigenvectors $\left|k,u;\dots\right>$ of the fourmomentum
operators\footnote{Notice that we have contravariant as well as
covariant fourmomenta related by $g_{\mu\nu}$; the physical
energy and momentum are covariant because they are generators of
translations.} namely
\begin{equation}
P^{\mu}\left|k,u;\dots\right>=k^{\mu}\left|k,u;\dots\right>        \label{11}
\end{equation}
where
\begin{equation}
\left<k',u;\dots|k,u;\dots\right>=2k_{+}^{0}\delta^{3}(\nad{k'}-\nad{k})
                                                               \label{12}
\end{equation}
i.e. we adopt a covariant normalization. The
$k_{+}^{0}=g^{0\mu}k^{+}_{\mu}$ is positive and $k^{+}_{0}$ is
the corresponding solution of the dispersion relation
\begin{equation}
k^2\equiv g^{\mu\nu}k_{\mu}k_{\nu}=-\kappa^2.                  \label{13}
\end{equation}
Namely
\begin{equation}
k_{0+}=-\frac{\vec{\sigma}}{c}\nad{k}+\gamma_{0}^{2}\omega_{k}
\label{14}
\end{equation}
with
\begin{equation}
\omega_k=\gamma_{0}^{-2}\sqrt{\left(\frac{\vec{\sigma}\nad{k}}{c}\right)^2+
\left(|\nad{k}|^2-\kappa^2\right)\gamma_{0}^{2}}.
\label{15}
\end{equation}
Notice that $k_{+}^{0}=\omega_{k}$ and the range of the
covariant momentum $\nad{k}$ is determined by the following inequality
\begin{equation}
|\nad{k}|\geq\kappa\left(1+\left(\gamma_{0}^{2}-1\right)
\left(\frac{\vec{\sigma}\nad{k}}
{|\vec{\sigma}||\nad{k}|}\right)^2\right)^{-1/2},
\label{16}
\end{equation}
i.e.\ values of $\nad{k}$ lie outside the oblate spheroid with halfaxes
$a=\kappa$ and $b=\kappa\gamma^{-1}_{0}$.
The covariant normalization
in (\ref{12}) is possible bacause in CT synchronization the sign
of $k^0$ is an invariant. Thus we have no problem with an
indefinite norm in $H_u$.

Now, $ku\equiv k_{\mu}u^{\mu}$ is an additional invariant.
Indeed, because the transformations of
$L_S$ are restricted to $SO(3)_u$ subgroup by causality
requirement, and $SO(3)_u$ does not change $u$ nor $k$, our
covariance group reduces\footnote{In fact $SO(3)_u$
acts like an $SO(3)$
intrinsic symmetry!} to the Poincar\'e mappings (realized in the
CT synchrony). Summarizing, irreducible family of unitary
operators $U(\Lambda,a)$ in the bundle of Hilbert spaces $H_u$
acts on an orbit defined by the following covariant conditions
\begin{itemize}
\item
$k^2=-\kappa^2$ ;
\item
$\varepsilon(k^0)=\mbox{\em inv}$; for physical
representations $\varepsilon(k^0)=1$ which guarantee a
covariant lower bound of energy \cite{Rem3} .
\item
$q\equiv\frac{uk}{c}=\mbox{\em inv}$; it is easy to see that $q$ is
an energy of tachyon measured in the privileged frame.
\end{itemize}
As a consequence there exists an invariant, positive definite measure
\begin{equation}
d\mu(k,\kappa,q)=d^4\!k\theta(k^0)\delta(k^2+\kappa^2)\delta(q-\frac{uk}{c})
\label{17}
\end{equation}
in a Hilbert space of wave packets.

Let us return to the problem of classification of irreducible
unitary mappings $U(\Lambda,a)$:
\[
U(\Lambda,a)\left|k,u;\dots\right>=\left|k',u';\dots\right>;
\]
here the pair $(k,u)$ is transported along trajectories
belonging to an orbit fixed by the above mentioned invariant
conditions. To follow the familiar Wigner procedure of induction
one should find a stability group of the double $(k,u)$. To do
this, let us transform $(k,u)$ to the prefered frame by the
Lorentz boost $L_{u}^{-1}$. Next, in the privileged frame, we
rotate the spatial part of the fourmomentum to the $z$-axis by
an appriopriate rotation $R^{-1}_{\vec{n}}$. As a result, we
obtain the pair $(k,u)$ transformed to the pair
$(\Nad{k},\Nad{u})$ with
\begin{equation}
\Nad{k}=\left(\begin{array}{c}
q\\   0\\   0\\   \sqrt{\kappa^2+q^2}
\end{array}\right),\qquad
\Nad{u}=\left(\begin{array}{c}
c\\  0\\  0\\  0
\end{array}\right).                       \label{18}
\end{equation}
It is easy to see that the stability group of
$(\Nad{k},\Nad{u})$ is the $SO(2)=SO(2,1)\cap SO(3)$ group.
Thus tachyonic unitary representations should be induced from
the $SO(2)$ instead of $SO(2,1)$ group! Recall that unitary
representations of the $SO(2,1)$ noncompact group are
infinite dimensional (except of the trivial one). As a consequence, local
fields was necessarily infinite component ones (except of the
scalar one). On the other hand, in the CT synchronization case
unitary representations for space-like fourmomenta in our bundle
of Hilbert spaces are induced from irreducible, one dimensional
representations of $SO(2)$ in a close analogy with a light-like
fourmomentum case. They are labelled by helicity
$\lambda$, by $\kappa$ and by $q$
($\varepsilon(k^0)=\varepsilon(q)$ is determined by $q$; of
course a physical choice is $\varepsilon(q)=1$).

Now, by means of the familiar Wigner trick we determine the
Lorentz group action on the base vectors; namely
\begin{equation}
U(\Lambda)\left|k,u;\kappa,\lambda,q\right>=
e^{i\lambda\varphi(\Lambda,k,u)} \left|k',u';\kappa,\lambda,q\right>
                                             \label{19}
\end{equation}
where
\begin{equation}
e^{i\lambda\varphi(\Lambda,k,u)}=U\left(R^{-1}_{\Omega\vec{n}}
\Omega R_{\vec{n}}\right)                           \label{20}
\end{equation}
with
\begin{equation}
\Omega=L^{-1}_{u'}\Lambda L_u.                \label{21}
\end{equation}
Here $k$ and $u$ transform according to the law (\ref{1}). The
rotation $R_{\vec{n}}$ connects $\Nad{k}$ with
$D(L^{-1}_{u},u)k$, i.e.
\begin{equation}
R_{\vec{n}}\Nad{k}=D(L^{-1}_{u},u)k.            \label{22}
\end{equation}
It is easy to check that $R^{-1}_{\Omega\vec{n}}\Omega
R_{\vec{n}}$ is a Wigner-like rotation belonging to the
stability group $SO(2)$ of $(\Nad{k},\Nad{u})$ and
determines the phase $\varphi$. By means of standard topological
arguments $\lambda$ can take integer or halfinteger values only
i.e. $\lambda=0, \pm 1/2, \pm 1, \dots.$

 Now, the orthogonality
relation (\ref{12}) reads
\begin{equation}
\left<k',u;\kappa',\lambda',q'|k,u;\kappa,\lambda,q\right>=
2\omega_k\delta^{3}(\nad{k'}-\nad{k})\delta_{\lambda',\lambda}.
                                                               \label{23}
\end{equation}


\subsection{Canonical quantization}
Following the Fock procedure, we define canonical commutation
relations
\begin{mathletters}
\begin{equation}\label{24}
[a_{\lambda}(k_+,u),a_{\tau}(p_+,u)]_{\pm}=
[a_{\lambda}^{\dagger}(k_+,u),a_{\tau}^{\dagger}(p_+,u)]_{\pm}=0,
\end{equation}\begin{equation}
\mbox{}[a_{\lambda}(k_+,u),a_{\tau}^{\dagger}(p_+,u)]_{\pm}=
2\omega_k\delta(\nad{k}-\nad{p})\delta_{\lambda\tau},
\end{equation}
\end{mathletters}
where $-$ or $+$ means the commutator or anticommutator and
corresponds to the bosonic ($\lambda$ integer) or fermionic
($\lambda$ halfinteger) case respectively. Furthermore, we
introduce a Poincar\'e invariant vacuum $\left|0\right>$ defined
by
\begin{equation}\label{25}
\left<0|0\right>=1 \qquad\mbox{and}\qquad
a_{\lambda}(k_+,u)\left|0\right>=0.
\end{equation}
Therefore the one particle states
\begin{equation}
a_{\lambda}^{\dagger}(k_+,u)\left|0\right>                        \label{26}
\end{equation}
are the base vectors belonging to an orbit in our bundle of
Hilbert spaces iff
\begin{mathletters}\label{27}
\begin{equation}
U(\Lambda)a_{\lambda}^{\dagger}(k_+,u)U(\Lambda^{-1})=
e^{i\lambda\varphi(\Lambda,k,u)}
a_{\lambda}^{\dagger}(k_{+}^{\prime},u^{\prime}),\end{equation}
\begin{equation}
U(\Lambda)a_{\lambda}(k_+,u)U(\Lambda^{-1})=
e^{-i\lambda\varphi(\Lambda,k,u)}a_{\lambda}(k_{+}^{\prime},u^{\prime}),
\end{equation}\end{mathletters}
and
\begin{equation}\label{28}
[P_{\mu},a_{\lambda}^{\dagger}(k_+,u)]_{-}=
k_{\mu}^{+}\,a_{\lambda}^{\dagger}(k_+,u).
\end{equation}
Notice that
\begin{equation}
P_{\mu}=\int d^4\!k\,\theta(k^0)\,\delta(k^2+\kappa^2)\,k_{\mu}\left(
\sum_{\lambda}a_{\lambda}^{\dagger}(k,u)a_{\lambda}(k,u)\right)
                                                                \label{29}
\end{equation}
is a solution of (\ref{28}).

Finally we can deduce also the form of the helicity operator:
\begin{equation}
\hat{\lambda}=-\frac{W^{\mu}u_{\mu}}
{\displaystyle c\sqrt{\left(\frac{Pu}{c}\right)^2 -P^2}}
  \label{30}
\end{equation}
where
\[
W^{\mu}=\frac{1}{2}\varepsilon^{\mu\sigma\lambda\tau}J_{\sigma\lambda}
P_{\tau}
\]
is the Pauli-Lubanski fourvector.

The next step is to construct local free fields and the corresponding
field equations. An example of such construction it was presented in
\cite{Rem3} for a free scalar tachyonic field. Here we give a local
field equation for Dirac tachyons; it can be possibly associated with
neutrinos
\begin{equation}
\left[-i\partial_{\mu}\gamma^\mu+\frac{1}{2\xi}\left(\kappa^2
+\left(i\frac{u^\mu}{c}\partial_\mu+\xi\right)^2\right)
\frac{u_\mu\gamma^\mu}{c}-\frac{1}{2\xi}\left(\kappa^2+
\left(i\frac{u^\mu}{c}\partial_\mu+\xi\right)\left(i\frac{u^\mu}{c}
\partial_\mu-\xi\right)\right)I\right]
\psi(x)=0                                                       \label{31}
\end{equation}
with $\xi$ is a real parameter of inverse of length dimensionality;
here $\psi(x)$ is a Dirac bispinor, whereas
\[
\gamma^{\mu}=T(u)^{\mu}_{\;\;\nu}\gamma^{\nu}_{E},
\]
i.e.,
\begin{equation}
\left\{\gamma^{\mu},\gamma^{\nu}\right\}=2g^{\mu\nu}(u)I        \label{32}
\end{equation}
{}From (\ref{31}) it follows the Klein-Gordon equation
\begin{equation}
\left(g^{\mu\nu}(u)\partial_{\mu}\partial_{\nu}-\kappa^2\right)\psi=0,
                                                                \label{33}
\end{equation}
so in the momentum representation we have the tachyonic dispersion relation
\begin{equation}
k^2=-\kappa^2. \label{34}
\end{equation}
Recall that in the standard approach it is impossible to introduce a
mass term without breaking of hermicity of the Lagrangian describing
fermionic tachyon.

Now, the helicity operator (\ref{30}) takes the form
\begin{equation} \label{35}
\hat{\lambda}=\frac{1}{\displaystyle
4c\sqrt{\left(\frac{Pu}{c}\right)^2-P^2}}
\gamma^5[P\gamma,u\gamma].
\end{equation}
Because $\hat{\lambda}^2=\frac{1}{4}I$ and $\mathop{\rm Tr}\hat{\lambda}=0$,
the spectrum of $\hat{\lambda}$ equals
$\{-\frac{1}{2},-\frac{1}{2},\frac{1}{2},\frac{1}{2}\}$. Furthermore
$\hat{\lambda}$ commutes with the Dirac operator defined by (\ref{31}).
This allows us to construct solutions of (\ref{31}) by means of standard
procedure, i.e., by using the projection operators constructed from
helicity and Dirac operators.

\section{Conclusions}
The main result of this work is that tachyons are classified according
to the unitary representations of $SO(2)$ rather than $SO(2,1)$ group.
Together with the fact that they can be consistently described under
some appropriate choice of synchronization, this shows that there are no
serious theoretical obstructions to interprete the experimental data
about square of mass of neutrinos as a signal that they can be fermionic
tachyons.

A more exhaustive discussion of the Dirac-like equation for
fermionic tachyon will be given in the forthcoming article.


\begin{thebibliography}{10}

\bibitem{tab}
{Particle Data Group}, Phys. Rev. {\bf D50},  1390  (1994).

\bibitem{CHK}
A. Chodos, A.~I. Hauser, and V.~A. Kostelecky, Phys. Lett. {\bf B150},  431
  (1985).

\bibitem{Kos1}
V.~A. Kostelecky, in {\em Topics on Quantum Gravity and Beyond},
edited by F. Mansouri and T.~T. Scanio
(World Scientific, 1993).

\bibitem{CKPG}
A. Chodos, V.~A. Kostelecky, R. Potting and E. Gates, Mod. Phys.
Lett. {\bf A7}, 467 (1992).

\bibitem{CK}
A. Chodos and V.~A. Kostelecky, Phys. Lett. {\bf B336}, 295 (1994).

\bibitem{Rem3}
J.Rembieli\'nski, 
Preprint KFT U{\L} 2/94, Katedra Fizyki Teoretycznej Uniwersytetu
{\L}{\'o}dzkiego, HEP-TH/9410079 (submitted to Phys. Rev. {\bf  D}).

\bibitem{Rem1}
J. Rembieli{\'n}ski, Phys. Lett. {\bf A78},  33  (1980).

\bibitem{Rem2}
J. Rembieli{\'n}ski, Preprint IF U{\L}/1/1980, Instytut Fizyki Uniwersytetu
  {\L}{\'o}dzkiego (unpublished).

\bibitem{Rei}
H. Reichenbach, {\em Axiomatization of the Theory of Relativity} (University of
  California Press, Berkeley, CA, 1969).

\bibitem{Jam}
M. Jammer,  in {\em Problems in the Foundations of Physics} (North-Holland,
  Bologne, 1979).

\bibitem{VT}
J.~G. Vargas and D.~G. Torr, Found.\ of Phys. {\bf 16},  1089  (1986).

\bibitem{Var}
J.~G. Vargas, Found.\ of Phys. {\bf 16},  1003  (1986).

\bibitem{Sjo}
T. Sj{\"o}din, Nuov.\ Cim. {\bf B51},  229  (1979).

\bibitem{Cha1}
T. Chang, Phys. Lett. {\bf A70},  1  (1979).

\bibitem{Cha2}
T. Chang, J. Phys. {\bf A13},  L207  (1980).

\bibitem{Cha3}
T. Chang, J. Phys. {\bf A12},  L203  (1979).

\bibitem{Tan}
F.~R. Tangherlini, Nuov.\ Cim. Suppl. {\bf 20},  1  (1961).

\bibitem{Sud}
E.~C.~G. Sudarshan, in {\em Tachyons, Monopoles and Related Topics},
  edited by E. Recami (North-Holland, New York, 1978).

\end{thebibliography}

\end{document}